\documentclass[twocolumn, prd, amssymb, preprintnumbers]{revtex4}
\usepackage{amssymb}
\usepackage{amsmath}

\usepackage{graphicx}

\newcommand{\bea}{\begin{eqnarray}}
\newcommand{\eea}{\end{eqnarray}}
\newcommand{\be}{\begin{equation}}
\newcommand{\ee}{\end{equation}}

\begin{document}  

\preprint{ITFA 2009-18, arXiv:0907.2695v3}
\title{Holographic Neutron Stars}
\author{Jan de Boer, Kyriakos Papadodimas and Erik Verlinde}
\affiliation{ Institute for Theoretical Physics, University of Amsterdam\\
              Valckenierstraat 65, 1018 XE Amsterdam, The Netherlands\\
              J.deBoer, K.Papadodimas, E.P.Verlinde @uva.nl} 

\bigskip

\date{July 2009}

\begin{abstract}
We construct in the context of the AdS/CFT correspondence degenerate composite operators in the conformal field theory  that are holographically dual to degenerate stars in  anti de Sitter space.   
We calculate the effect of the gravitational back-reaction using the Tolman-Oppenheimer-Volkoff equations, and determine the "Chandrasekhar limit" beyond which the star undergoes gravitational collapse towards a black hole.

\end{abstract}

\maketitle

\section{Introduction}

A lot of progress has been made  in string theory in understanding the most extreme form of matter, namely black holes. String theory should also be able to describe less extreme forms of matter, such as the degenerate matter that makes up neutron stars. 
A neutron star is able  to withstand the force of gravity  as long as the total mass stays below a critical value referred to as the  Oppenheimer-Volkoff limit \cite{Oppenheimer}.   This critical mass is determined by solving the Einstein equations for spherically symmetric pressure and energy density  distributions, also known as the Tolman-Oppenheimer-Volkoff equations.  

 It would be especially  interesting to study the collapse of a neutron star towards a black hole in a context with a holographic description.  In this letter we take a first step in this direction by studying degenerate stars in Anti-de Sitter space, and constructing the corresponding states within the conformal field theory.  
 One of our aims is to determine the OV limit for these degenerate stars and to give its holographic interpretation in terms of the CFT. Previous studies of stars in AdS  are found in \cite{Page}.

We begin in section II by describing  degenerate composite operators in the CFT constructed out of many fermionic single trace primary fields. In section III we show that these operators correspond to a degenerate Fermi gas in AdS space, and we present a hydrodynamic description.  We subsequently define a double scaling limit, and argue that in this limit the mass and Fermi energy receive corrections due to self gravity only.  The TOV equations in AdS space and their analytic properties are described in section IV. In section V  we present our numerical results, compare with the boundary calculations and conclude with remarks on the interpretation.

\section{Degenerate conformal operators}

We consider a  CFT  defined on the cylinder $R\times S^{d-1}$   with metric
\be
\label{cylindermetric}
ds^2= -dt^2+\ell^2 d\Omega_{d-1}^2,
\ee
 where $d\Omega_{d-1}^2$ represents the standard metric on $S^{d-1}$. 
The cylinder is identified with the boundary of  global anti de Sitter space $AdS_{d+1}$ with curvature radius $\ell$. To make use of the operator state correspondence, 
we analytically continue to euclidean time and  conformally map the cylinder on the euclidean plane $R^d$ by 
$$
t\to i \ell \log|x|.
$$
The angular coordinates  $\Omega$ are identified in the obvious way.   
States in the CFT correspond to operators via
\be
\lim_{|x|\to 0} \Phi(x)\left | {\rm vac}\right \rangle= \left |\Phi\right\rangle.
\ee
We now assume that  the CFT  contains bosonic as well as fermionic fields, and has a large $c$ limit  in which one can distinguish single and  multi trace operators \footnote{Throughout this paper we use $c$ instead of the rank $N$. For  $SU(N)$ ${\cal N}\!=\!4$ SYM the relation is: $N^2\!-\!1=\pi^4 c/10$.}.

Our aim is to find a degenerate state made out of fermions. So let us choose a fermionic single trace primary operator $\Psi$ with conformal dimension $\Delta_0$.  $\Psi$ is assumed to be complex so that it differs from its conjugate $\overline{\Psi}$. 
We first consider the limit $c\to\infty$, while keeping all other quantities fixed. In this limit single trace primaries behave as "free" fields, since  three and higher point functions are suppressed by powers of $1/\sqrt{c}$, see footnote [5]. 
By spin-statistics, fermionic operators anti-commute
\be
\Psi(x)\Psi(y)=-\Psi(y)\Psi(x),
\ee
 and hence they generate a free fermionic Fock space.  The lowest dimension composites of two fields  $\Psi$ are given by
$\Psi\partial_i\Psi$, where $\partial_i$ denotes the derivative with respect to  $x_i$.   We continue with a third fermionic operator, a fourth, up to a very large number,  each time  making sure that the next operator is anti-symmetrized with respect to the previous ones by acting with one, two , three derivatives, etc.  For a degenerate state one uses  the smallest number of derivatives necessary to avoid that any given combination occurs twice. For simplicity we ignore spinor indices. In this way we arrive at the following composite operator 
\be
\begin{split}
\label{Phi}
{\bf \Phi}= \Psi \prod_i  \partial_i \Psi\prod_{\lbrace i,j \rbrace}\partial_i\partial_j \Psi \prod_{\lbrace i,j,k \rbrace}\partial_i\partial_j \partial_k \Psi & \ldots\!\!\!\!  \\ \ldots\ldots \prod_{\lbrace i_1, i_2,\ldots i_{n_F}\rbrace}\!\!\!\!\!\partial_{i_1}&\partial_{i_2} \ldots\partial_{i_{n_F}}\!\! \Psi
\end{split}
\ee
The fermionic primaries  are arranged in "shells" labeled by the number of derivatives, where in each shell the product is taken over all possible $n$-tuples of the derivatives $\partial_i$.   The number of derivatives in the last shell is denoted by $n_F$.  We consider the case with a completely filled last shell, so that we have a rotationally symmetric operator. 
The total number of fields contained in $\Phi$  may then be expressed in terms of binomial coefficients as 
\be
\label{NF}
N= \sum_{n=0}^{n_F}
\left(\! \begin{array}{c}\! n \! +\! d\!-\! 1\! \\ \! d\!-\! 1\! \end{array}\! \right).
\ee
Since each derivative carries conformal dimension one,  each operator in the $n$-th shell has dimension $n+\Delta_0$. Hence the total conformal dimension 
of  $\bf \Phi$ equals 
\be
\label{DeltaF}
\Delta = \sum_{n=0}^{n_F}(n+\Delta_0)
\left(\! \begin{array}{c}\! n \! +\! d\!-\! 1\! \\ \! d\!-\! 1\! \end{array}\! \right). 
\ee
The primary operator $\Psi$ corresponds to a fermionic  field $\psi$ in the bulk.  The number of derivatives acting on $\Psi$ is directly related to the quantum number $n$ that determines the energy of the partial waves of $\psi$, see e.g.\ \cite{AdSCFT}. The operator $\Phi$  represents a degenerate many particle state, whose complete wave function is given by the Slater determinant  of all partial waves with $n\leq n_F$.   

Let us now introduce the total mass $M$,  the fermion mass $m$,   and the Fermi energy $\epsilon_F$ via
$$
m=\Delta_0/\ell,\qquad M=\Delta/\ell, \qquad \epsilon_F=( n_F+\Delta_0)/\ell.
$$
We are interested in the limit  of large $n_F$ (and large $\Delta_0$), so it is more convenient to rewrite the sums in the expressions (\ref{NF}) and (\ref{DeltaF}) as integrals. This gives
\be
\label{relation}
N(\epsilon_F) = \int^{\epsilon_F}_m  \!\!\!\! d\epsilon \,g(\epsilon)\, ,\qquad M(\epsilon_F)  =\int^{\epsilon_F}_m  \!\!\!\! d\epsilon \,\, \epsilon\, g(\epsilon)\, ,
\ee
where the density of states $g(\epsilon)$ is given by
\be
\label{DOS}
g(\epsilon) = \frac{\ell^d}{(d\!-\!1)!}(\epsilon-m)^{d-1}
\ee
This expression is valid in the strict large $c$ limit.

In the following we define a double scaling limit that takes $n_F$  and $c$  simultaneously to infinity, while keeping the ratio $\Delta/c$ fixed. In this  limit the 't Hooft suppression partially breaks down, and therefore there will be  corrections to the conformal dimension $\Delta$ and density of states $g(\epsilon)$ defined in (\ref{relation}).  The main correction to $\Delta$ turns out to come from $t$-channel exchange of the stress tensor, and  can be estimated  to be of the order of $\Delta^2/c$, see \cite{toappear}.  This corresponds  to the inclusion of  self gravity in the bulk.

When operator mixing sets in,  the state $|\Phi\rangle$ has to be defined a little more carefully. Suppose we have a state $|\Phi\rangle_N$ composed out of $N$ operators $\Psi$. We then define 
\be
\Psi (x) |\Phi\rangle_{{}_{N}} \sim  x^{(m-\epsilon_{{}_F})\ell}\, |\Phi\rangle_{{}_{N+1}} ,
\ee
where one chooses $\epsilon_F$  so that one picks up (one of) the lowest order term(s) on the right hand side. For $\Delta/c \to 0$ this should lead to the operator (\ref{Phi}). This procedure is well defined as long as the density of states $g(\epsilon_F)$  remains finite.  When $g(\epsilon_F)$ diverges  the definition of $\Phi$ becomes obscure.  We will see this is precisely what happens at the Oppenheimer-Volkoff limit.

\section{Hydrodynamic Description}

We return to the case where first $c\to \infty$, and then take a  subsequent limit in which  the number of particles per AdS volume goes to infinity while keeping the ratio $\epsilon_F/m$ fixed.
In this hydrodynamic limit the particles form an ideal degenerate Fermi gas. The number of particles within one AdS radius is very large, and therefore the energy density $\rho$ and  pressure $p$ obey the  same equation of state as in flat space. Specifically, the particle number density $\bf n$  equals the volume of the Fermi sphere
\be
\label{number}
{\bf n} 
 = \frac {V_{d-1}}{d (2\pi)^d}\, k_F^d,
\ee
where $V_{d-1}$ denotes the area of  $S^{d-1}$. The Fermi momentum $k_F$ is related to the chemical potential $\mu$ via
\be
\label{ekf}
\mu= \sqrt{k_F^2+m^2}.
\ee
The energy density and pressure are given by  the "textbook" expressions
\bea
\label{textbook}
\rho &=& \frac {V_{d-1}}{(2\pi)^d}\int_0^{k_F}\!\!\!dk\,k^{d-1}  \sqrt{k^2+m^2} ,\nonumber \\
p & =&  \frac {V_{d-1}}{d (2\pi)^d}\int_0^{k_F}\!\!\!dk\,\frac{k^{d+1} }{ \sqrt{k^2+m^2}} ,
\eea
and obey the 
standard thermodynamic relations
\be
\label{identity}
 d\rho\   = \mu\, d {\bf n}, \qquad
\rho+p  = \mu\, {\bf n}. 
\ee
We want to determine the density and pressure profile corresponding to the operator $\Phi$. We will do so first for the case of pure $AdS$,  but in the following section we generalize these results to include self-gravity. Anticipating these generalizations, let us write the metric in the form
\be
\label{Ansatz}
ds^2 = - A(r)^2 dt^2 + {B(r)^2} {dr^2}+ r^2 d\Omega^2_{d-1}.
\ee
The stress energy tensor of the Fermi gas is given by the standard hydrodynamic expression
\be
T_{\mu\nu} = (\rho+p)u_{\mu} u_{\nu} + p g_{\mu\nu},
\ee
where $u_\mu$  is a static velocity field:  $u_\mu dx^\mu = A(r)dt$. 
The radial profiles of $p$ and $\rho$ are determined by imposing stress energy conservation. 
This leads to the condition
\be
\label{dpdr}
\frac{dp}{dr} +\frac{1}{A}\frac{dA}{dr} (\rho+p) =0.
\ee
This equation is surprisingly easy to solve. By making use of the identities  (\ref{identity})
one easily verifies that (\ref{dpdr}) is satisfied when the chemical potential obeys
\be
\label{redshift}
\mu(r)= \frac{\epsilon_F}{ A(r)},
\ee
where at this stage $\epsilon_F$ is an arbitrary constant.  We conclude that the radial dependence of the chemical potential  is simply due to the gravitational redshift. We now observe that the constant $\epsilon_F$ equals the chemical potential defined with respect to the time $t$. This means that it can be identified with the Fermi energy $\epsilon_F$ of the CFT. 
 
 When the chemical potential $\mu$ is equal to the fermion mass  $m$,   the pressure and energy density drop to zero. From (\ref{redshift}) we find that this happens at $r\!=\!R$ for which 
\be
\label{edge}
A(R)= {\epsilon_F\over m}.
\ee
This defines the edge of the star. Note that the edge is only present when the fermions have  a finite mass. 

Up to this point our discussion holds for any metric of the form (\ref{Ansatz}). We now specialize to the case of pure AdS space, so that 
$A^2= 1/B^{2}=1+{r^2/ \ell^2}$.  One then finds that the radius $R$ of a free Fermi gas in AdS is  determined by the Fermi momentum at the center:
$
R/\ell={ k_F(0)/ m}.
$

As a non-trivial check on our hydrodynamic description one can verify that the particle number and mass agree with the boundary calculations. 
The total particle number is given by the integral
\be
\label{integralN}
N= V_{d-1}  \int_0^R\!\!  dr \, r^{d-1}{B(r)}\,{{\bf n}(r)}.
\ee
The factor $B(r)$ comes from the volume form on the spatial section. 
Similarly, one obtains the mass $M$  by integrating the energy density. Because $M$ is defined with respect to the time coordinate $t$ one has to include a redshift factor $A(r)$. For AdS this cancels the measure factor $B(r)$, and hence
\be
\label{integralM}
M= V_{d-1}  \int_0^R\!\!  dr \,r^{d-1}\rho(r).\,
\ee
In fact,  this expression  is equivalent to the ADM mass, and also holds when self gravity is included.
To calculate $N$ and $M$ one has to insert the equations (\ref{number}) and (\ref{textbook}) for the particle and energy density and include the redshift effect (\ref{redshift}) into the chemical potential.  The resulting integrals can be performed analytically and (perhaps not) surprisingly precisely reproduce the results (\ref{relation}) and (\ref{DOS}).

\section{Including Self Gravity}

Having established that for infinite $c$ the operator $\Phi$  is represented by a free degenerate Fermi gas, we now proceed to the double scaling limit $n_F \to \infty$ and $c\to \infty$  with fixed values for 
$$
\frac{n_F}{\Delta_0}=\frac{\epsilon_F}{m}-1, \qquad
\frac{\Delta}{c} \sim {GM\over\ \ell^{d-2}}.
$$ 
One may think this leads not only to gravitational interactions, but also possibly makes the fermions unstable. However, the 't Hooft suppression still seems to be sufficiently active to ensure that the Fermi gas has an infinite life time.  The possible  depletion of the particle number density is controlled by the
Boltzman equation
\be
\frac{d{\bf n}}{dt}\sim{k_F\over m}\,{\bf n}^2\, \sigma ,
\ee
where $\sigma$ represents the total cross section associated with the decay.  In the double scaling limit  $\bf n$ grows like $c^{d/(d\!+\!1)}$, while $\sigma$ contains 
at least one factor of $c^{-1}$ due to 't Hooft suppression. So, unless $\sigma$ contains other factors of $c$, the life time of the fermions  is estimated to be of the order of $c^{1/ (d\!+\!1)}$, hence is infinite for $c\to \infty$. 
The decay rates  are possibly enhanced by energy or phase space factors. Various checks, involving dimensional arguments,  softness of string amplitudes and the optical theorem, indicate that these effects do not ruin our argument. For a  more detailed discussion we refer to \cite{toappear}. 

So let us now assume that gravity is the only force  \footnote{For  chiral primaries there is also a gauge force due to the $R$-charge. One finds that the results are qualitatively the same as presented here: gravity eventually wins \cite{toappear}.}.  
Except for including gravity, the fermionic matter will be treated exactly as in the previous section.  We parametrize the functions $A(r)$ and $B(r)$ in terms of two new functions $M(r)$ and $\chi(r)$ as
\bea
{A^2(r)} &= &e^{2\chi(r)}\left(1-\frac{C_d M(r)\,}{\ \ r^{d-2}}+\frac{r^2}{\ell^2}\right),\nonumber \\
{B^2(r)} &= & \left(1-\frac{C_d M(r)\,}{\ \ r^{d-2}}+\frac{r^2}{\ell^2}\right)^{\!\! -1}\!\!\!,\label{definitions}
\eea
where Newton's constant is contained in the coefficient
$$
C_d =\frac{16\pi \,G\ }{(d\!-\!1)V_{d-1} }.
$$
 In  terms of  $M(r)$ and $\chi(r)$ the Einstein equations read
\bea
M'(r) &=  & \, V_{d-1} \,\rho(r)\, r^{d-1} ,  \nonumber\\
 \chi'(r)& =&  V_{d-1}  \,\frac{C_d }{2}{\Bigl(\rho(r)+p(r)\Bigr)} \, r{B^2(r)} .
\label{einstein}
\eea
Together with (\ref{dpdr}) these constitute the TOV equations. We choose as boundary condition $M(0)\!=\!0$, so that $M(r)$  represents the contribution to the mass from the energy density inside a ball of radius $r$, and the total mass is equal to $M(R)$.
By Birkhoff's theorem the metric  outside the star is given by AdS-Schwarzschild.  Hence, 
$$
\chi(r)=0, \qquad M(r)=M,\qquad  \mbox{ for $r\geq R$}.
$$
From (\ref{edge}) it follows that the radius $R$  is determined by 
\be
1-\frac{C_d M\,}{\ \ R^{d-2}}+\frac{R^2}{\ell^2}=\left({\epsilon_F\over m}\right)^2.
\ee
We want to use the TOV equations to calculate the corrections to the mass, the particle number and the density of states defined through the relations  (\ref{relation}). In fact, to show that  the relations (\ref{relation}) are still valid  one needs to use  the Einstein equations!  One has
\bea
\mbox{${}$}\!\!\! \epsilon_F\frac{dN}{d\epsilon_F}&\!=\!& V_{d\!-\!1} \int^R_0 \!\!\! dr \, r^{d-1} \left( B \, \epsilon_F \frac{d {\bf n}}{d \epsilon_F}+   \frac{d B}{d \epsilon_F}\epsilon_F{\bf n}\right) \nonumber\\
&\!=\!& V_{d\!-\!1} \int^R_0 \!\!\! dr \,  r^{d-1} e^{\chi} \left(\,  \frac{d \rho}{d \epsilon_F} +   \frac{d B}{d \epsilon_F} \frac{\rho+p}{B}\right) \nonumber \\
&\!=\!&  \int^R_0 \!\! dr  \, e^{\chi }  \left(\frac{d M'}{d \epsilon_F}+ \frac{dM}{d\epsilon_F}{\chi'}\right)=\frac{dM}{d\epsilon_F}.
\eea
Here we used the identity (\ref{identity}), the relation (\ref{redshift}), the TOV equations (\ref{einstein}),  and the definitions of $A$ and $B$ in terms of the functions $M$ and $\chi$.

\begin{figure}[tbp]
\begin{center}
\includegraphics[scale=0.38]{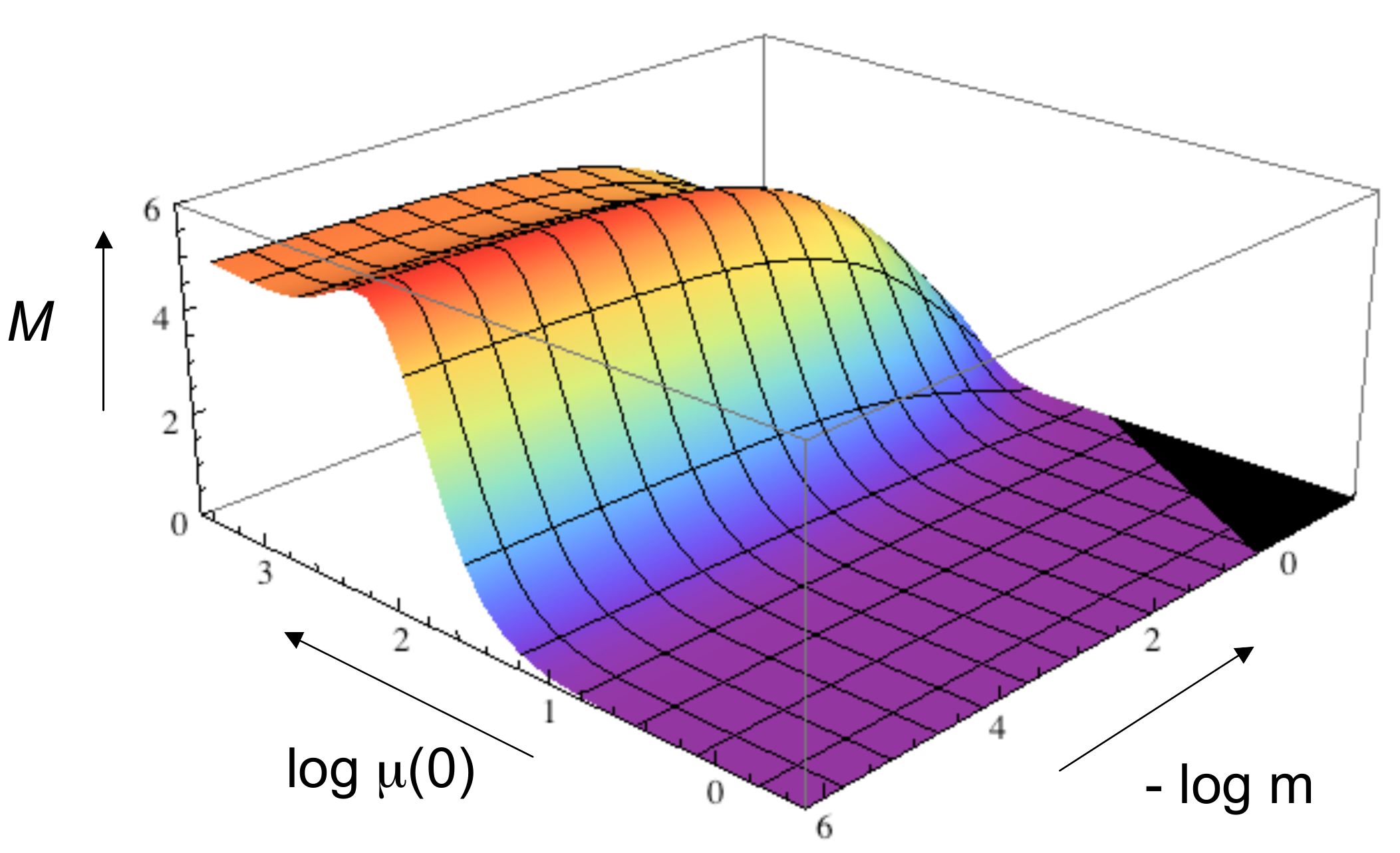}
\end{center}
\caption{$M$ as a function of $\log \mu(0)$ plotted against $\log m$ for  $AdS_5$. The black corner is the "forbidden" region $\mu(0)<m$.}
\end{figure}
\begin{figure}[tbp]
\begin{center}
\includegraphics[scale=0.38]{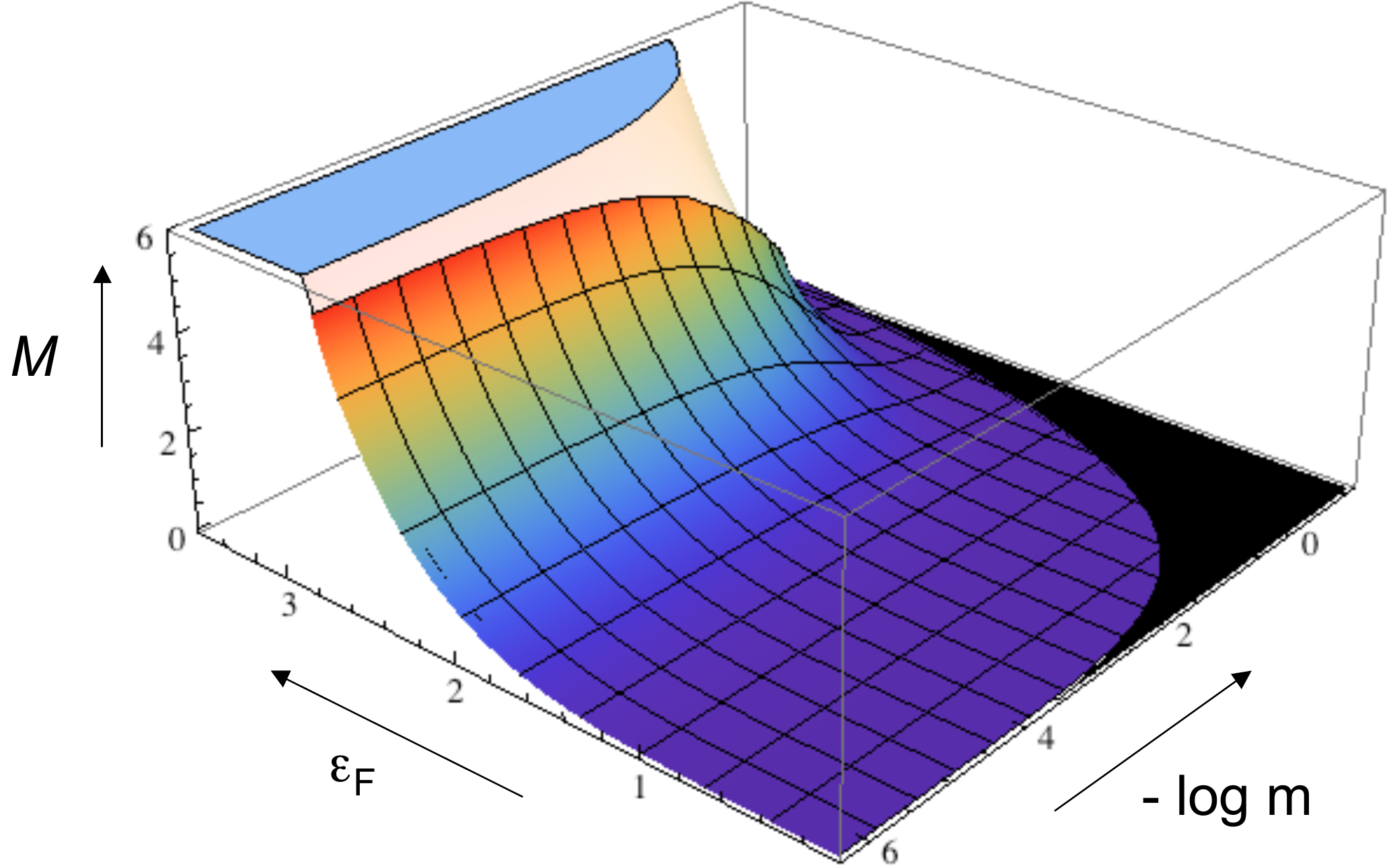}
\end{center}
\caption{$M(\epsilon_F)$ plotted against $\log m$.  The underlying pink graph is the result without gravity. The units in all figures are  $M=\Delta/c$, $\epsilon_F=(n_F\!+\!\Delta_0)/c^{1/(d\!+\!1)}$ and $m=\Delta_0/c^{1/(d\!+\!1)}.$}
\end{figure}

\begin{figure}[tbp]
\begin{center}
\bigskip
\includegraphics[scale=0.38]{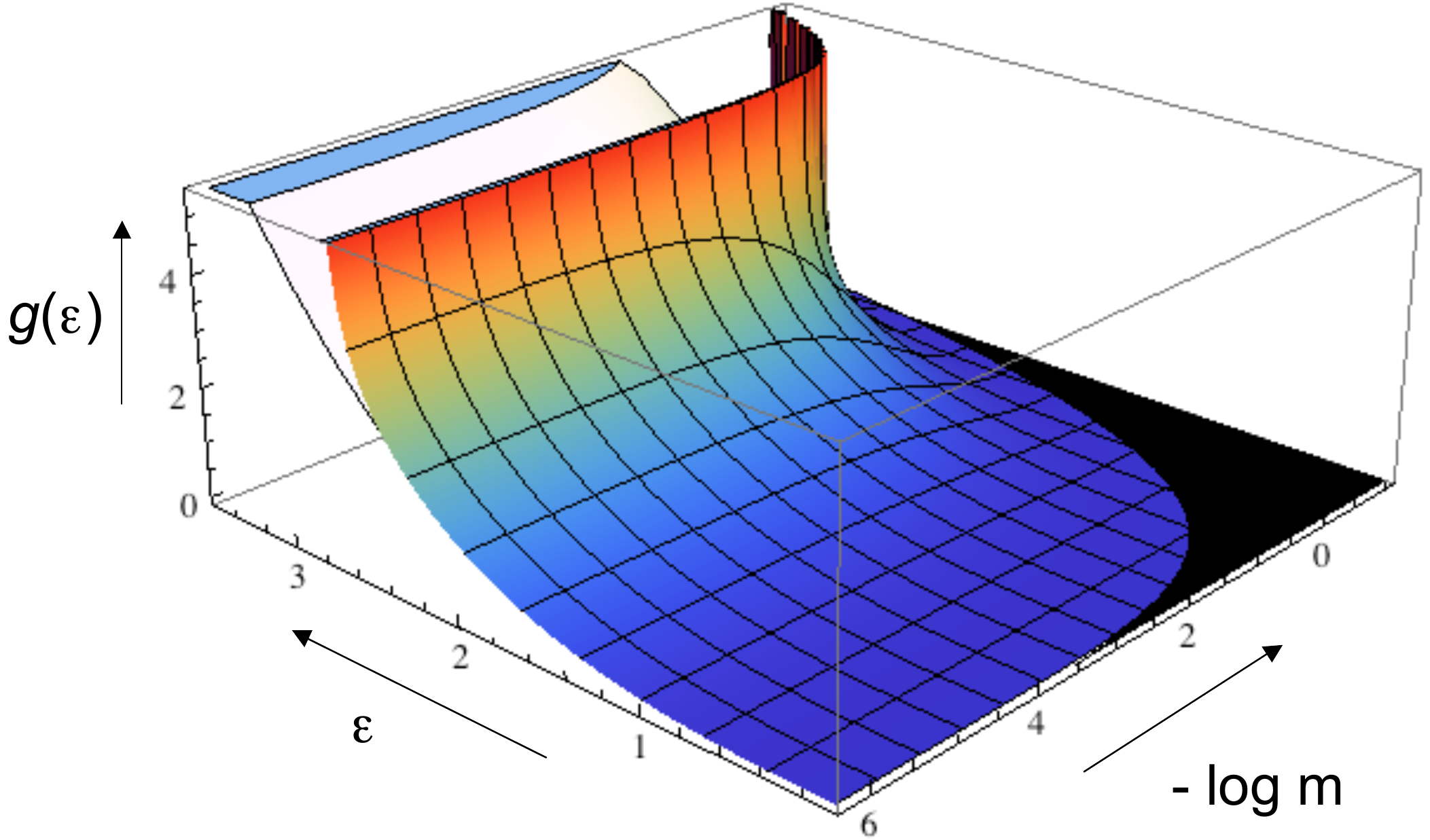}
\end{center}
\caption{The density of states $g(\epsilon)$  plotted against $\log m$. The underlying graph is the result (\ref{DOS}) without back-reaction. Here and in FIG.2 the black region corresponds to $\epsilon_F<m$.}
\end{figure}

\section{Results and discussion.}

The TOV equations can be integrated numerically. We give the results for $d=4$, so for gravity in $AdS_5$.  Other cases are qualitatively similar. Just as in \cite{Oppenheimer, Page} we find that the mass $M$ reaches a maximum as a function of  the chemical potential at the center, see FIG 1. Beyond this critical value the solution becomes unphysical. Similar plots are found for  $N$ and $\epsilon_F$ as a function of $\mu(0)$.  From a holographic viewpoint it is more natural to regard $M$ as a function of the Fermi energy $\epsilon_F$.  This leads to the graph shown in FIG 2. The upper edge of the colored part is the OV limit, while the underlying graph is the un-backreacted result (\ref{relation}) and (\ref{DOS}).    The effect of self gravity on the value of $M$ is surprisingly small. The same holds for $N$. Yet on the density of states defined as $g(\epsilon)\!=\!dN/d\epsilon$ the effect is significant: it diverges at the OV limit as  shown in FIG. 3. This is a clear sign of an instability.  

Of course, one would like to derive the corrections to the mass  from the boundary perspective. Qualitatively it is clear that these corrections are due to operator mixing.  More importantly, we like to find the meaning of  the OV limit in the  CFT. In the bulk the star becomes unstable under a radial density perturbation. It would be interesting to interpret this mode in the CFT. Beyond the OV limit the star presumably collapses and forms a black hole.  In the CFT this means that the operator $\Phi$ starts to mix with generic states, and thermalizes. This suggests that the collapse is associated with a phase transition that turns a high density (baryonic) state into a thermal state (quark gluon plasma). Clearly these questions and the stability issue need further study.

\medskip

\centerline{\bf Acknowledgments}
We would like to thank O. Aharony, S. Hellerman,  G. Horowitz, 
V. Hubeny, E. Kiritsis, J. Maldacena, P. McFadden,  S. Minwalla, B.
Nienhuis, H. Ooguri, M. Rangamani, L. Rastelli, K. Schoutens, M.
Shigemori, A. Strominger, M. Taylor, S. Trivedi, and H. Verlinde for discussions. 
This work was partly supported by the Foundation of Fundamental
Research on Matter (FOM).

\nopagebreak

\nopagebreak

\end{document}